\documentclass[aps,prl,groupedaddress]{revtex4-2}
\usepackage{chemformula} 
\usepackage{color,soul}
\usepackage{threeparttable}
\usepackage{multirow}
\usepackage{booktabs}
\usepackage{graphicx}
\usepackage[table,xcdraw]{xcolor}
\usepackage{url}
\usepackage{latexsym}
\usepackage{array}
\usepackage{bm}
\usepackage{subfigure}
\usepackage{amsfonts}

\begin{document}

\title{Large-gap quantum anomalous Hall insulator in two-dimensional pentagonal TiN\(_{8}\) monolayer}


\author{Zheng Wang$^{1,\#}$, Linda Sophie Reitz$^{4,\#}$, Beichen Ruan$^{1}$, Zhuoheng Li$^{1}$, Shu-Shen Lyu$^{1-3}$, Richard Dronskowski$^{4,*}$, Kaixuan Chen$^{1-3,*}$}

\email{chenkx26@mail.sysu.edu.cn; drons@HAL9000.ac.rwth-aachen.de}

\affiliation
{$^1$School of Materials, Shenzhen Campus of Sun Yat-sen University, Shenzhen, 518107, PR China, $^2$Guangdong Engineering Technology Research Centre for Advanced Thermal Control Material and System Integration (ATCMSI), Sun Yat-sen University, Guangzhou, 510275, PR China, $^3$Huizhou Research Institute, Sun Yat-sen University, Huizhou, 516081, PR China, $^4$Chair of Solid-State and Quantum Chemistry, Institute of Inorganic Chemistry, RWTH Aachen University, D-52056 Aachen, Germany. $^{\#}$These authors contribute equally to this work. $^{*}$Corresponding authors: Richard Dronskowski(drons@HAL9000.ac.rwth-aachen.de), Kaixuan Chen(chenkx26@mail.sysu.edu.cn)}


\date{\today}

\begin{abstract}

Discovering the quantum anomalous Hall (QAH) effect in two-dimensional magnets is a central goal of topological condensed matter physics, yet candidate materials combining a sizable gap with a well-defined Chern number remain scarce. We report a new material platform that realizes emergent magnetism and nontrivial band topology within the largely overlooked family of two-dimensional MN$_8$ monolayers. Through first-principles calculations, we reveal that these systems exhibit out-of-plane magnetic ground states and nontrivial topological features driven by the localized $d$-orbitals of the embedded transition metals. Remarkably, we identify TiN$_8$ as a quantum anomalous Hall insulator characterized by a Chern number of $C=-1$ with a large topological gap of 223 meV. As a companion system, MoN$_8$ is predicted to be a rare high-Chern-number QAH insulator, with a Chern number of $C=2$, although its much smaller gap of 12 meV renders it more fragile. These findings establish the penta-MN$_8$ family as a versatile and promising platform to realize exotic topological quantum states. This work not only broadens the material landscape for magnetic topological insulators but also lays a solid theoretical foundation for the development of next-generation spintronic and quantum computing devices.

\end{abstract}


\maketitle

\section{Introduction}
\noindent 

Recently, topological insulators (TI) have attracted significant attention as an untried frontier in condensed matter physics and materials science due to their unique electronic structure, characterized by an insulating bulk band gap and topologically protected gapless edge states\cite{mazumder2021brief,hasanColloquiumTopologicalInsulators2010, hasanThreeDimensionalTopologicalInsulators2011,denner2021exceptional}. Among them, two-dimensional (2D) Chern insulators, a class of intrinsic magnetic topological materials, exhibit a nonzero Chern number and support the quantum anomalous Hall effect, offering a promising platform for low-power electronic devices and topological quantum computing\cite{mazumder2021brief,hasanColloquiumTopologicalInsulators2010, hasanThreeDimensionalTopologicalInsulators2011,denner2021exceptional}. However, experimentally realized 2D magnetic TIs, such as Cr-doped (Bi,Sb)$_2$Te$_3$ thin films, face challenges regarding their synthesis and stability, which significantly limit their practical applications\cite{jiangAntiferromagneticChernInsulators2018, changExperimentalObservationQuantum2013a, dengQuantumAnomalousHall2020}.

As recently reported, the rapid development of two-dimensional van-der-Waals materials offers a promising path to overcome the current challenges faced by conventional magnetic TIs\cite{kang2024evidence,garcia2020canted,kaneQuantumSpinHall2005, zhangTopologicalInsulatorsBi2Se32009, liuQuantumSpinHall2011,li2024weyl,zhong5dOrbitalInduced2025}. These materials display a wide variety of topological phases and, thanks to their weak interlayer interactions and tunable electronic structures, promise excellent integration compatibility\cite{xueAntiferromagneticQuantumSpin2023,zhu2022high,bosnar2023high,yan2024exchange,wang2025effects}. As a new 2D family, pentagonal structures including penta-graphene and penta-MX$_2$, have been theoretically predicted and experimentally synthesized, demonstrating remarkable mechanical and electronic properties\cite{zhang2015penta,oyedele2017pdse2,bykov2021realization,nazir2022research,naseriInvestigationStabilityElectronic2018}. The recent discovery of the nitrogen-based macrocycle N$_{18}$ further enriches this family\cite{eremets2004single,duPentaMN8FamilyFirst2023b}. Subsequent first-principles studies predicted that, by embedding transition metal elements, the N$_{18}$ framework can form more stable penta-MN$_8$ structures, exhibiting intriguing properties such as flat bands and promising acoustic, optical and thermoelectric behavior\cite{duPentaMN8FamilyFirst2023b}. By focusing solely on the crystal structure and general physical properties, previous studies have overlooked the localized nature of $d$-orbitals in transition metals, thereby ignoring the resultant emergent magnetism and nontrivial band topology.

In this work, we perform first-principles spin-polarized calculations to determine the magnetic ground states of penta-MN$_{8}$ systems. Our results reveal that these compounds, once synthesized, should exhibit ferromagnetic ground states with an out-of-plane magnetic moment. The appearance of band crossings contributed by the $d$-orbitals of M atoms at the Fermi level gives rise to nonzero Berry curvatures and nontrivial band topology. Remarkably, TiN$_{8}$ is predicted to exhibit a Chern number of $C = -1$, while MoN$_{8}$ should show a higher Chern number of $C = 2$. This predictive discovery not only expands the family of two-dimensional magnetic topological materials but also provides a solid theoretical foundation for exploring strong correlation effects, nontrivial topological states, and their potential applications in electronic and spintronic devices.

\section{Results and discussion}
\subsection{Structural properties and stability}

The hexagonal MN$_8$ network originates from the N$_{18}$ macrocycle found in extreme pressure compounds of yttrium synthesized in 2022\cite{aslandukov2022anionic}. Given that structural idea, MN$_8$ results from the formal insertion of a transition metal (M) atom into the N$_{18}$ macrocycle, resulting in a distorted pentagonal motif by one M and four connected N atoms inside a hexagonal crystal structure\cite{duPentaMN8FamilyFirst2023b}. Its structural stability was confirmed by DFT calculations, suggesting that MN$_{8}$ compounds could potentially be synthesized, possibly under high temperature, most certainly using extreme pressure conditions. The MN$_8$ family could be categorized into four structural patterns, but in this study, we focus on two representatives: TiN$_8$, a $\gamma$-phase structure defined as having a purely planar geometry, which was previously predicted to be unstable\cite{duPentaMN8FamilyFirst2023b}; and MoN$_{8}$ , a $\beta$-phase structure characterized by a puckered morphology, where metal atoms lie in a central layer flanked by two nitrogen layers, as defined in previous work\cite{duPentaMN8FamilyFirst2023b}. 

The previous study reported a lattice parameter of $a=5.28$ Å for MoN$_8$ and noted that TiN$_8$ exhibits imaginary frequencies in its phonon dispersion\cite{duPentaMN8FamilyFirst2023b}, indicating dynamic instability. All these results, however, were obtained using simplified non-magnetic (spin-degenerate) DFT calculations, which neglect the possible spin polarization of 3$d$ transition metal elements.

\begin{figure*} [htbp]
    \centering
    \includegraphics[width=1\textwidth]{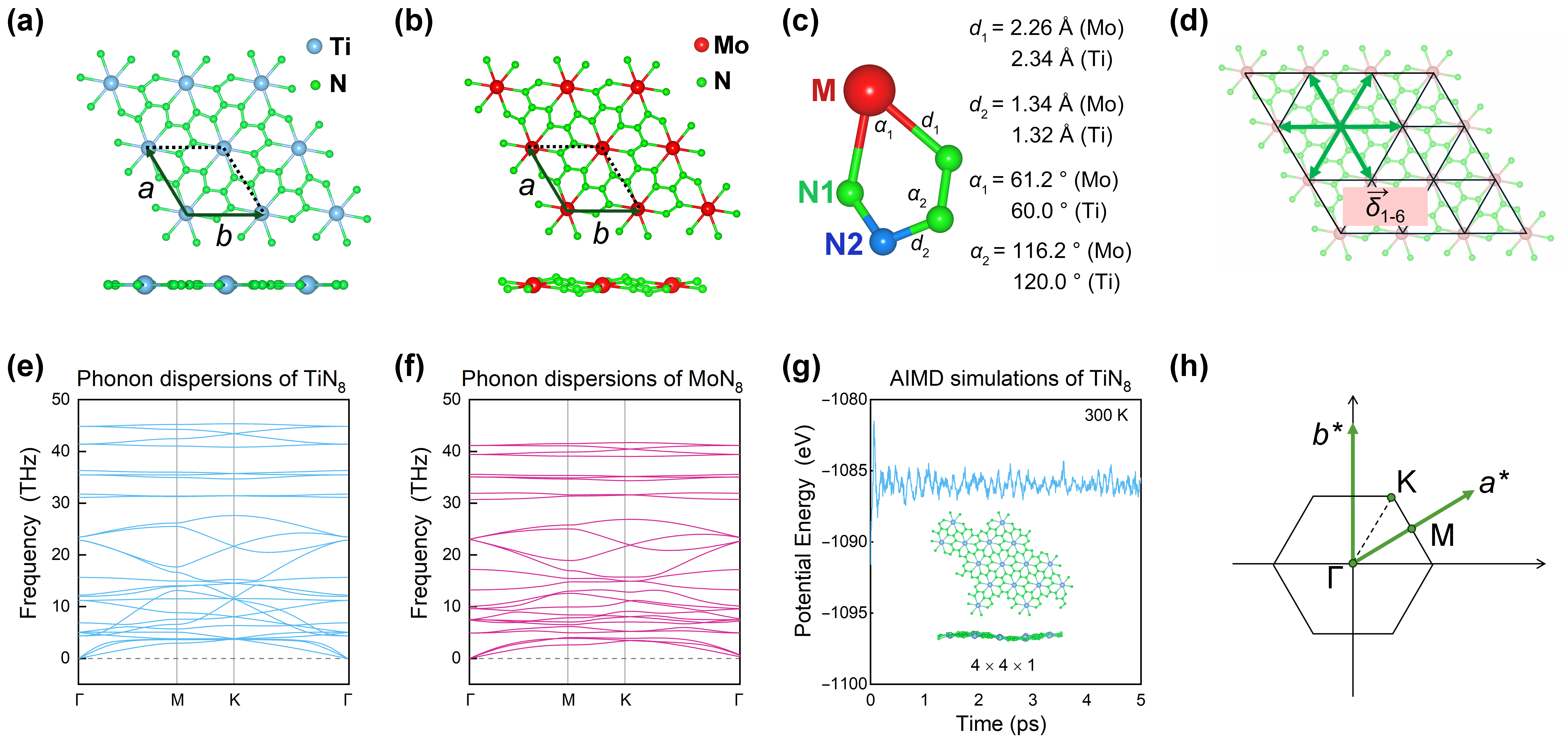}
    \caption{(a, b) Top and side views of the optimized TiN$_8$ and MoN$_8$ crystal structures, respectively. (c) Pentagonal motif showing bond lengths and angles for TiN$_8$/MoN$_8$. red, green, and blue spheres represent the M, N1, and N2 sites, respectively. (d) The hexagonal structural network composed of M atoms in a supercell, the six nearest neighbor sites are denoted as green arrows. (e),(f) The phonon dispersions of TiN$_8$/MoN$_8$, respectively. (g) evolution of the potential energy for a 4 × 4 × 1 supercell during the AIMD simulations at 300 K and a snapshot of the equilibrium structure (inset) at the end of 5 ps. (h) The first Brillouin zone of the MN$_8$ systems. }
    \label{fig1}
\end{figure*}

The re-calculated results reveal that MoN$_8$ and TiN$_8$ show hexagonal symmetry and lattice parameters of $a=5.47\,\mathrm{\mathring{A}}$ and $a=5.67\,\mathrm{\mathring{A}}$ respectively, indicating a slight increase in size upon introducing spin polarization, a result of the exchange hole. The results of phonon calculations and \textit{ab initio} molecular dynamics (AIMD) simulations are presented in Fig.\ref{fig1}(e, f, g), with both structures exhibiting structural stability to some extent. For TiN$_8$, the potential energy fluctuates around a stable average at 300\,K indicating thermal stability while MoN$_8$ fails to maintain structural integrity within 3\,ps, even when the temperature is lowered to 50\,K (as shown in Fig.S2). Energetically, MN$_8$ behaves similarly to nitrogen-rich compounds such as lead azide Pb(N$_3$)$_2$ = PbN$_6$, which, despite being above the convex hull by a formation enthalpy of +436.5\,kJ/mol, can be handled due to its metastability and sufficient activation barriers against decay. Upon considering the decomposition reaction 
\begin{equation*}
    \mathrm{MN}_8 \xrightarrow{} \mathrm{M(bulk)} + 4\mathrm{N}_{2}
\end{equation*}

 the calculated formation energies are +967\,kJ/mol and +902\,kJ/mol for TiN$_8$ and MoN$_8$ respectively. Given that cubic TiN and hexagonal MoN define the convex hull, with formation enthalpies of about -378 and -105 kJ/mol (reference to Materials Project), decomposition of the MN$_8$ to MN and N$_2$ will release only -589 and -797 kJ/mol, for the case of TiN$_8$ comparable to lead azide. This indicates that both systems are thermodynamically unstable, too, but may turn out as being metastable provided unusual conditions for synthesis, in particular for TiN$_8$.

Despite failing to maintain structural integrity for 3\,ps the intriguing mechanism of the high Chern number observed in MoN$_8$ still warrants its inclusion in our study, for sheer reasons of scientific curiosity. The structural details are presented in Fig.\ref{fig1}(a, b, c). The monolayer TiN$_8$ is predicted to crystallize in space group $P6/m$(175), while in-plane distortions in MoN$_8$ reduce its symmetry to $P\overline{3}$(147). Fig.\ref{fig1}(c) illustrates the unique pentagonal substructure in the MN$_8$ systems. The eight nitrogen atoms within a primitive cell can be categorized into two distinct groups. The first group, comprising six nitrogen atoms that are directly bonded circumferentially to the central metal (M) atom, is referred to as N1 sites and denoted in green. The second group consists of two nitrogen atoms that do not bond directly with the M atom but instead form bonds with the three neighboring N1 atoms. These are designated as N2 sites and are colored in blue. Fig.\ref{fig1}(d) presents the hexagonal structure composed of M atoms and the corresponding nearest neighbor sites.

\subsection{Electronic structure and magnetic properties}

\begin{figure} [htbp]
    \centering
    \includegraphics[width=0.5\textwidth]{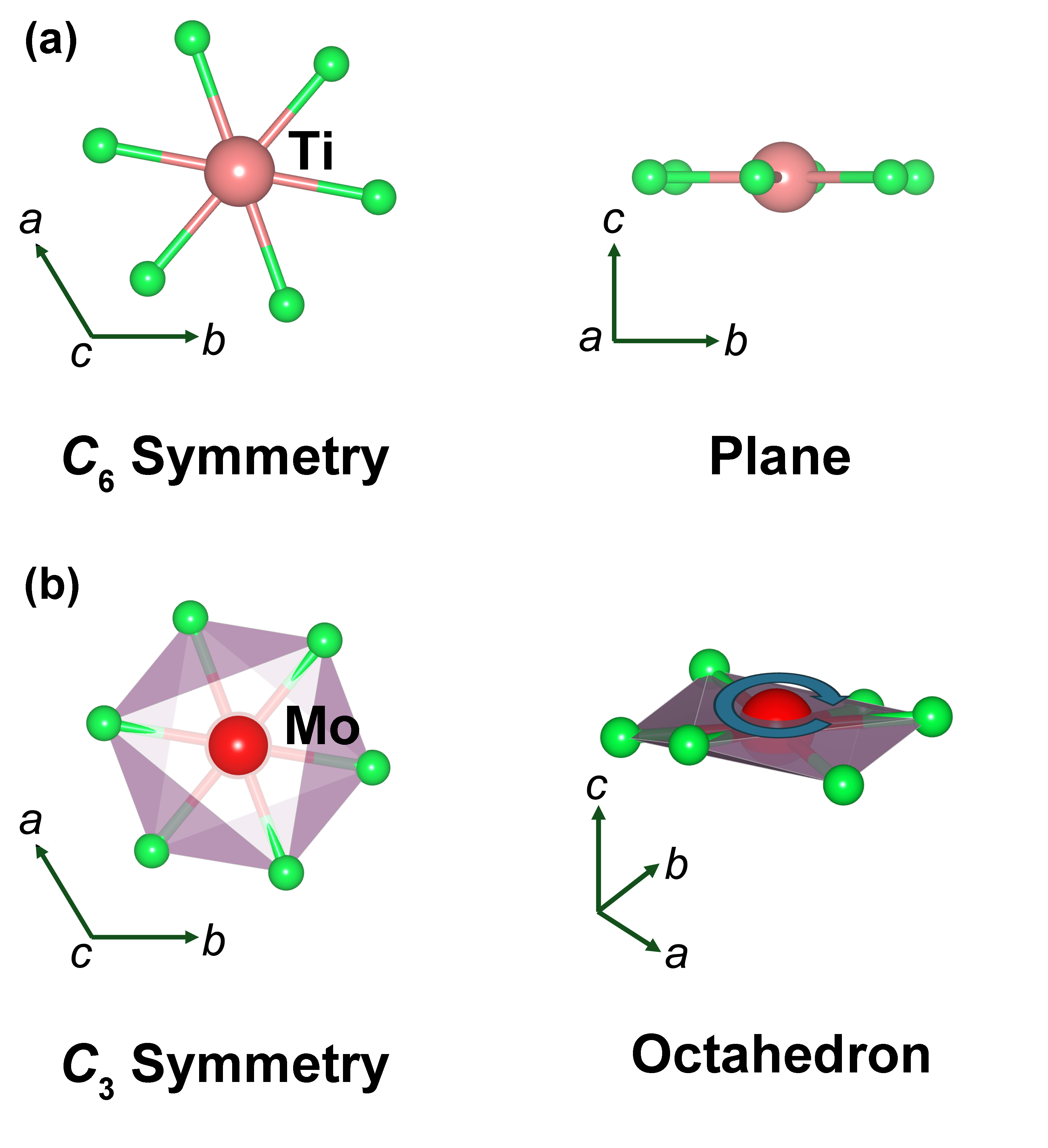}
    \caption{Ligand coordination environments of TiN$_8$ (a) and MoN$_8$ (b), respectively.}
    \label{fig2}
\end{figure}

\begin{figure} [htbp]
    \centering
    \includegraphics[width=0.5\textwidth]{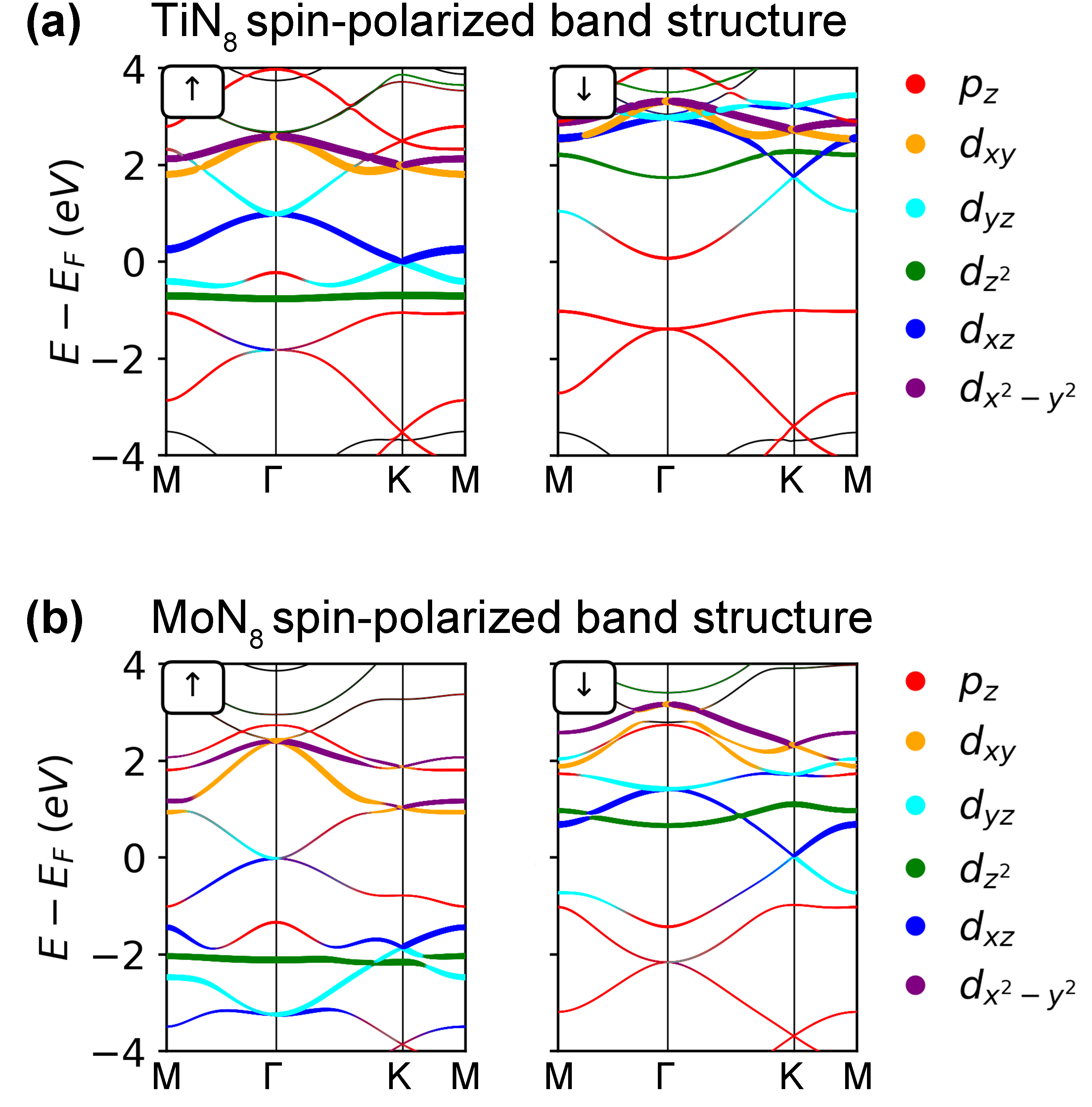}
    \caption{(a) and (b) Orbital-projected band structures of TiN$_{8}$ and MoN$_{8}$ for the spin-up and spin-down subspaces, respectively.}
    \label{fig3}
\end{figure}

The initially expected electronic structure of MN$_8$ is that of substantial polar-covalent interaction between the M and N atoms, so some ionic contribution, in addition to massive covalency between the N atoms. As shown in Fig.\ref{fig2}(a), TiN$_8$ adopts an atomically planar structure, where the N1 sites symmetrically surround the central Ti atom. However, in-plane puckering of MoN$_8$ introduces a compressed octahedral environment, resulting in a distorted crystal field as shown in Fig.\ref{fig2}(b). This distortion raises the level of $d{z^2}$ and significantly lowers $d_{xy}$, as indicated in Fig.\ref{fig3}(b).  Moreover, the space groups of TiN$_8$ ($P6/m$) and MoN$_8$ ($P\overline{3}$) preserve the $C_3$ rotational symmetry, which requires band degeneracies at the high-symmetry $\Gamma$ and $K$ points in the Brillouin zone. 

Fig.\ref{fig3}(a) and (b) show the orbital-projected band structures of MoN$_8$ and TiN$_8$ without spin-orbit coupling (SOC), respectively. Evidently, both structures exhibit pronounced spin polarization, with electronic features strongly influenced by crystal field shapes.  As can be observed from the fatband of TiN$_8$ in Fig.\ref{fig3}(a), the band $d{z^2}$ lies lowest in energy, followed by the bands ($d_{xz}$, $d_{yz}$), and finally ($d_{xy}$, $d_{x^2-y^2}$). The $d$ energy bands of MoN$_8$ resemble those of TiN$_8$, but are more closely related to an octahedral arrangement. The $p_z$ orbitals of the N atoms exhibit orbital mixing with the Ti $d_{yz}$ orbitals near the $\Gamma$ point. 

\begin{figure*} [htbp]
    \centering
    \includegraphics[width=1\textwidth]{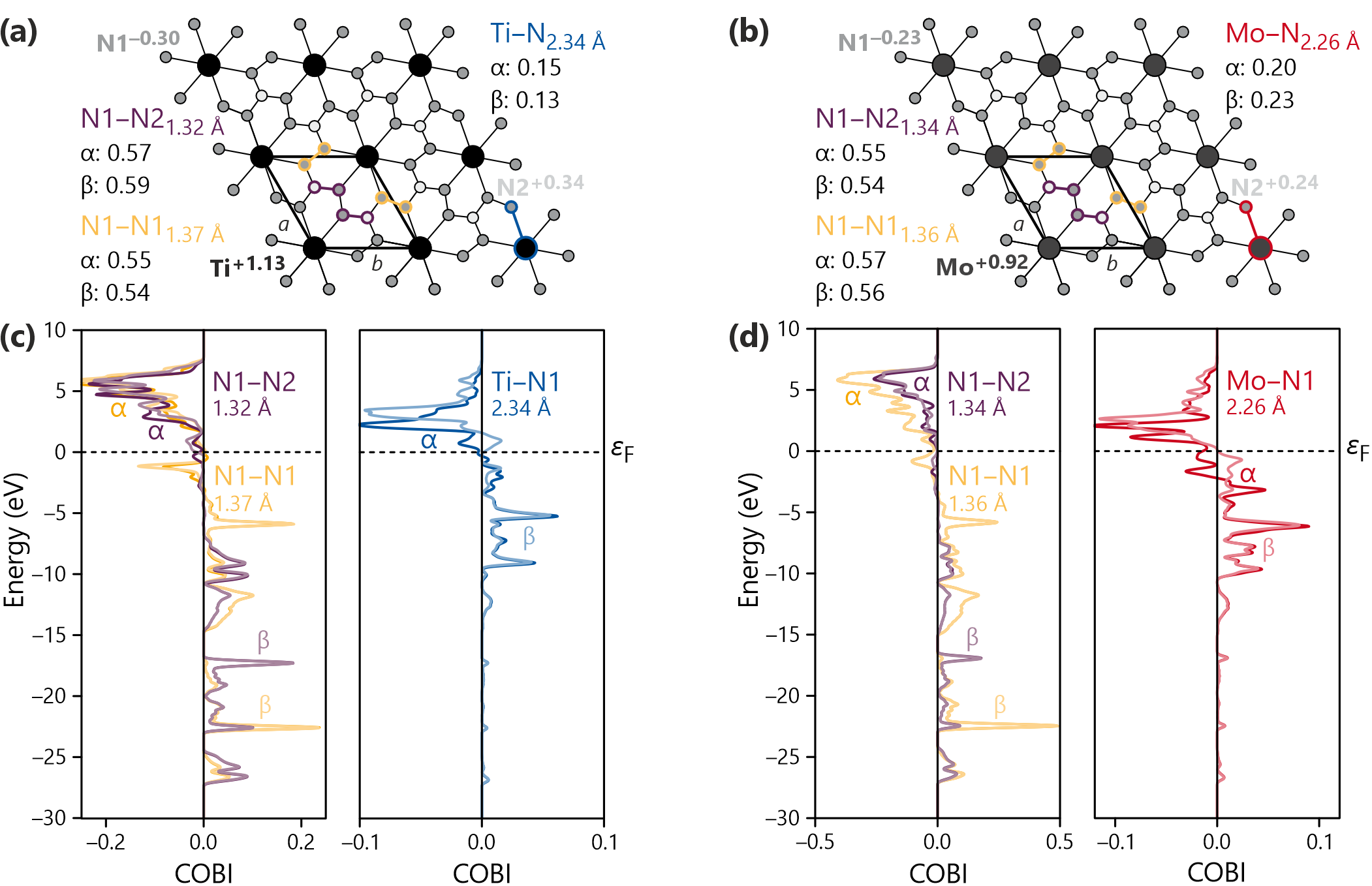}
    \caption{(a) and (b) Löwdin Charges and integrated Crystal Orbital Bond Index (ICOBI) for TiN$_8$ and MoN$_8$. (c) and (d) The Crystal Orbital Bond Index (COBI) for the two types of N--N bonds is shown in purple and yellow, while the Ti--N and Mo--N bonds are given in blue and red, respectively.}
    \label{fig4}
\end{figure*}

To shed some light on the bonding situation in the MN$_8$ systems, we performed a crystal orbital bond index (COBI) analysis, as shown in Fig.~\ref{fig4}(c, d). The COBI plot for the Ti--N bonds in TiN$_8$ (Fig.\ref{fig4}(c)) shows no anti-bonding contributions (which would appear on the left side of the plot) below the Fermi level. The integrated COBI (ICOBI) can be understood as the Lewis chemical bond order and is summarized in Fig.\ref{fig4}(a). The values obtained for the Ti-N bonds ($\alpha$ = 0.15, $\beta$ = 0.13, summing up to 0.28, a bond order of almost 1/3) confirm the significant covalent contributions. Similar observations can be made for MoN$_8$, given in Fig.\ref{fig4}(b). The Mo–N bonds are slightly shorter than the Ti--N bonds and exhibit even more covalency as expressed by the higher ICOBI values ($\alpha$ = 0.20, $\beta$ = 0.23, summing up to 0.43, almost a half bond). For the Mo-N bonds, a small anti-bonding contribution is found below the Fermi level around 0 to $-$3\,eV in the spin-up channel (the lower energy states are labeled as spin-up channel), so the two surplus electrons of Mo compared to Ti show up here. 

\begin{figure}[htbp]
    \centering
    \includegraphics[width=0.5\textwidth]{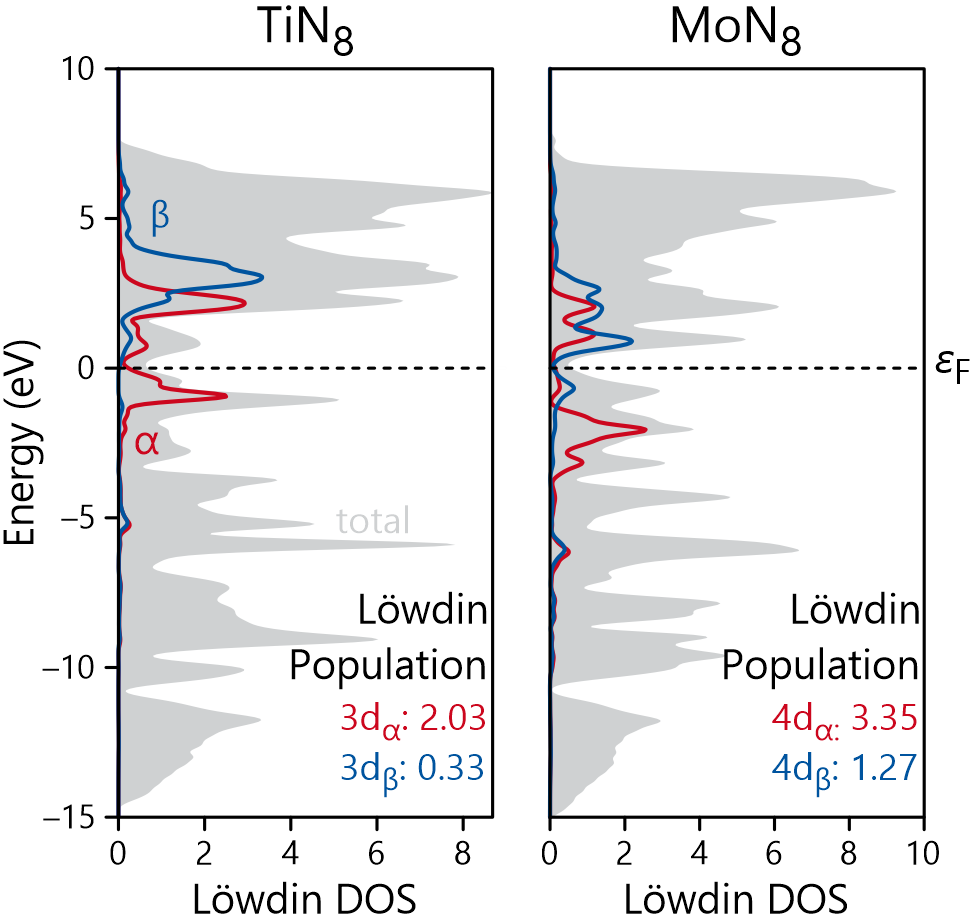}
    \caption{Spin-resolved Löwdin DOS for TiN$_8$ (left) and MoN$_8$ (right). The population of the d-orbitals is given for each spin channel.}
    \label{fig5}
\end{figure}

An analysis of the Löwdin charges reveals substantial charge transfer between the M atoms and their surrounding N1 atoms. In TiN$_8$ the charge transfer is greater
(Ti$^{+1.13}$, N1$^{-0.30}$) than in MoN$_8$ (Mo$^{+0.92}$, N1$^{-0.23}$), which makes TiN$_8$ slightly more ionic than MoN$_8$, in harmony with the electronegativities. The N2 atoms that do not bond directly with the M atom exhibit positive, cation-like Löwdin charges in both systems. In this way, there is also charge transfer between the N2 atoms and the N1 atoms with which they bond. These bonds can be identified as covalent bonds (including a polar contribution due to the different charge), and by summing up the ICOBI values for both spin channels, we can see that their bond order is slightly higher than one ($\sum \text{ICOBI}^\text{TiN$_8$}_\text{N1-N2} = 1.16$, $\sum \text{ICOBI}^\text{MoN$_8$}_\text{N1-N2} = 1.06$). This placement between a double and a single bond is also in line with what we would expect simply by looking at the bond lengths. The N--N bond lengths found in both systems range from 1.32 Å to 1.37 Å, so right between an N=N double (1.25 Å) and an N--N single bond (1.47 Å)\cite{zumdahl2016chemistry}. These interactions, combined with the presence of unpaired $d$ electrons, give rise to the observed magnetic moments of 1.92 and 2.00 $\mu_B$ per primitive cell for TiN$_8$ and MoN$_8$, respectively. The magnetic moment arrangements are simply generalized as $3d^2\uparrow$ and $4d^3\uparrow d^1\downarrow$ for TiN$_8$ and MoN$_8$ respectively (exact values are given alongside the DOS depicted in Fig.\ref{fig5}) despite the orbital mixing. 

 To describe the exchange coupling interaction, we conduct a 2D Heisenberg model Hamiltonian which is described as:

\begin{equation}
    H=H_0 -\sum_{\langle i,j\rangle}J_1S_i\cdot S_j-\sum_{\langle\langle i,j\rangle\rangle}J_2S_i\cdot S_j-\sum_iAS_i^2
\end{equation}

Here, $\langle i,j \rangle$ and $\langle\langle i,j \rangle\rangle$ denote the nearest-neighbor and second-nearest-neighbor pairs, respectively. The spin magnitude $S$ is set to 1 for both Ti and Mo atoms, corresponding to the calculated magnetic moments of approximately 2$\mu_B$. The parameters $J_1$ and $J_2$ represent isotropic exchange interactions, while $A$ denotes the magnetic anisotropy energy (MAE), calculated as the energy difference between an arbitrary spin orientation ($E_{xyz} $) and the out-of-plane spin orientation perpendicular to the $xy$-plane ($ E_{001}$). 

\begin{figure*} [htbp]
    \centering
    \includegraphics[width=\textwidth]{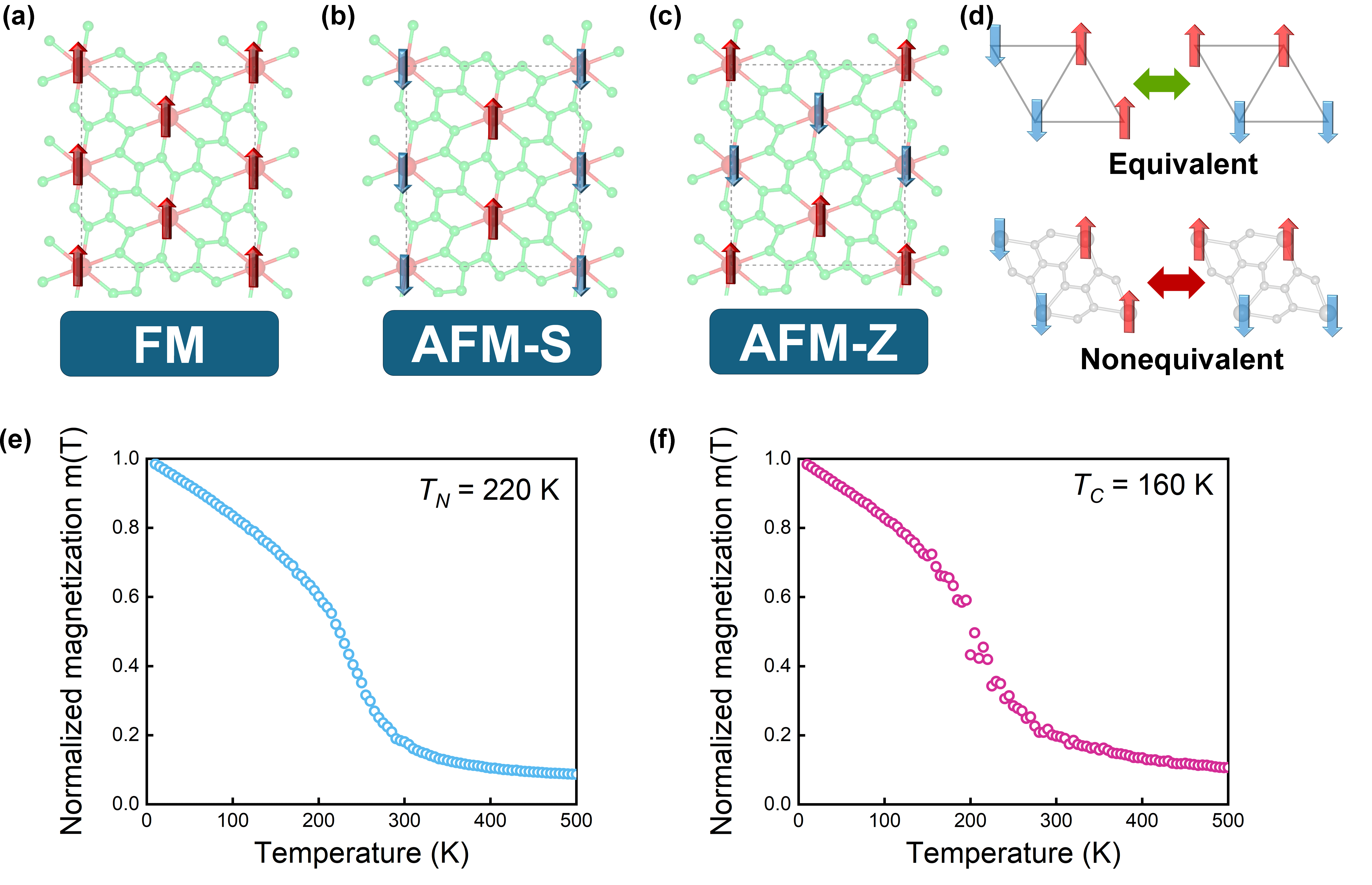}
    \caption{Exchange interactions. (a-c) Three different types of magnetic configurations implemented in the square lattice, which are (a) FM, (b) AFM-stripe and (c) AFM-zigzag, respectively. (d) Chirality in the triangular lattice and MN$_8$ lattice. (e, f) Curie/N\'{e}el of TiN$_8$ and MoN$_8$, respectively. }
    \label{fig6}
\end{figure*}

Since each MN$_8$ primitive cell contains only one magnetic site, magnetic exchange interactions must be investigated by constructing an extended supercell. However, the complex arrangement of nitrogen sites within the primitive cell breaks the mirror symmetry, resulting in a chiral characteristic as illustrated in Fig.\ref{fig6}(d). Our calculations show that non-equivalent magnetic configurations exhibit negligible energy differences (9.65 KJ/mol), allowing us to reasonably neglect the effects of the chiral N sites on the magnetic interactions as the exchange splitting on N is insignificant. Consequently, we simplified the system to a magnetic-site-only lattice depicted in Fig.\ref{fig6}(d), where magnetic configurations can be classified as FM or AFM states within a $2\times2\times1$ supercell. Based on this simplified structure, we further employ a redefined square supercell to model and analyze various spin configurations, including FM, AFM-stripe (AFM-S), and AFM-zigzag (AFM-Z) states, illustrated in Fig.\ref{fig6}(a, b, c). The total energies of these magnetic configurations are listed in Table S1 of the Supplemental Material, and the derived magnetic parameters are summarized in Table 1. The Curie/N\'{e}el temperatures ($T_{C}/T_{N}$) were estimated by Monte Carlo simulations using moderate Hubbard $U$ values of 3\,eV and 2\,eV for Ti and Mo respectively.

\begin{table*}[]
\caption{Exchange constants and magnetocrystalline anisotropy of MN$_8$ compared using different functionals.}
\label{tab:my-table}
\centering
\small
\begin{tabular}{ccccccc}
\hline\hline
structure                 & \multicolumn{3}{c}{TiN$_8$}  & \multicolumn{3}{c}{MoN$_8$} \\ \hline
Magnetism related constants & $J_{\text{1}}$      & $J_{\text{2}}$      & $A$     & $J_{\text{1}}$  & $J_{\text{2}}$ & $A$     \\ \hline
PBE+$U$  &-25.85 meV& 0.57 meV& 50.4 $\mu$eV & 46.27 meV&-14.41 meV&    1.60 meV\\
PBE   & 30.85 meV & -3.41 meV & 40.7 $\mu$eV& 11.43 meV&3.22 meV& 2.18 meV\\
SCAN  & 42.29 meV & -4.31 meV & 39.6 $\mu$eV& 21.53 meV & -7.76 meV& 1.82 meV\\ \hline\hline
\end{tabular}%

\end{table*}

As shown in Table 1, using various exchange functionals, both TiN$_8$ and MoN$_8$ prefer an out-of-plane easy axis for the spin orientation, where TiN$_8$ exhibits a magnetic anisotropy energy (MAE) on the order of $\mu$eV, for MoN$_8$ the order is meV. Regarding the exchange constants, both systems are dominated by the nearest-neighbor exchange constant $J_1$, which is larger than the second-nearest-neighbor exchange constant $J_2$. For TiN$_8$, $J_1$ is negative in the PBE + $U$ framework, indicating AFM exchange, while the PBE and SCAN calculations yield a positive $J_1$, suggesting the FM state. This highlights the strong $U$-dependence of the exchange interaction and questions, admittedly, the first-principles approach as everything depends on $U$. In MoN$_8$, the magnitude of $J_1$ is approximately three times that of $J_2$, indicating a robust FM state.

The Hubbard $U$ parameters were varied in the range of 1 to 4 eV to explore spin-polarized ground states. As shown in Table 2 and Fig.S1, TiN$_8$ adopts an AFM ground state when $U > 2.5$ eV, while MoN$_8$ consistently stabilizes in the FM state across the full range of $U$ values considered in this work. This behavior could be attributed to the competition between the localization of $d$ electrons and the Stoner itinerant electron effect. For TiN$_8$, when the Coulomb repulsion on-site $U < 2.5$, the Stoner effect dominates the magnetic exchange coupling; however, as $U > 2.5$, the Ti--N--N--Ti path which exhibits larger bond angles allows the hopping between the Ti $d$ and N $p_z$ orbitals, enhancing the superexchange-like effect. However, for MoN$_8$, the presence of more $d$ orbital electrons near the Fermi level promotes stronger itinerancy, enhancing the Stoner effect. Furthermore, the Mo--N--N--Mo pathway exhibits smaller bond angles, which would disfavor an AFM superexchange contribution; the shorter Mo-Mo distance also contributes to stronger direct exchange interactions or increased metallic bonding. These factors collectively ensure that the Stoner itinerant electron effect prevails largely throughout the range of $U$ values, leading to a preference for the FM state. The Stoner itinerancy is also mirrored from Fig.\ref{fig4}(d) as there are occupied antibonding Mo--N levels at the Fermi level, a chemical fingerprint of ferromagnetism.\cite{Landrum1999}

\begin{table*}[]
\caption{Energy differences $\Delta E$ (defined as $E_{AFM}-E_{FM}$) for both systems within 2$\times$2$\times$1 supercell}
\label{tab:my-table2}
\centering
\small
\begin{tabular}{cccccccc}
\hline\hline
Hubbard $U$ for Ti/Mo (eV)      & 1.0    & 1.5    & 2.0    & 2.5    & 3.0    & 3.5    & 4.0     \\ \hline
$\Delta E$ in TiN$_8$ (meV) & 158.94 & 101.13 & 24.48  & -20.36 & -56.92 & -87.27 & -117.19 \\
$\Delta E$ in MoN$_8$ (meV) & 118.96 & 136.51 & 146.45 & 148.27 & 151.34 & 146.54 & 153.84  \\ \hline\hline
\end{tabular}%

\end{table*}

\subsection{Topological properties}

\begin{figure*} [htbp]
    \centering
    \includegraphics[width=1\textwidth]{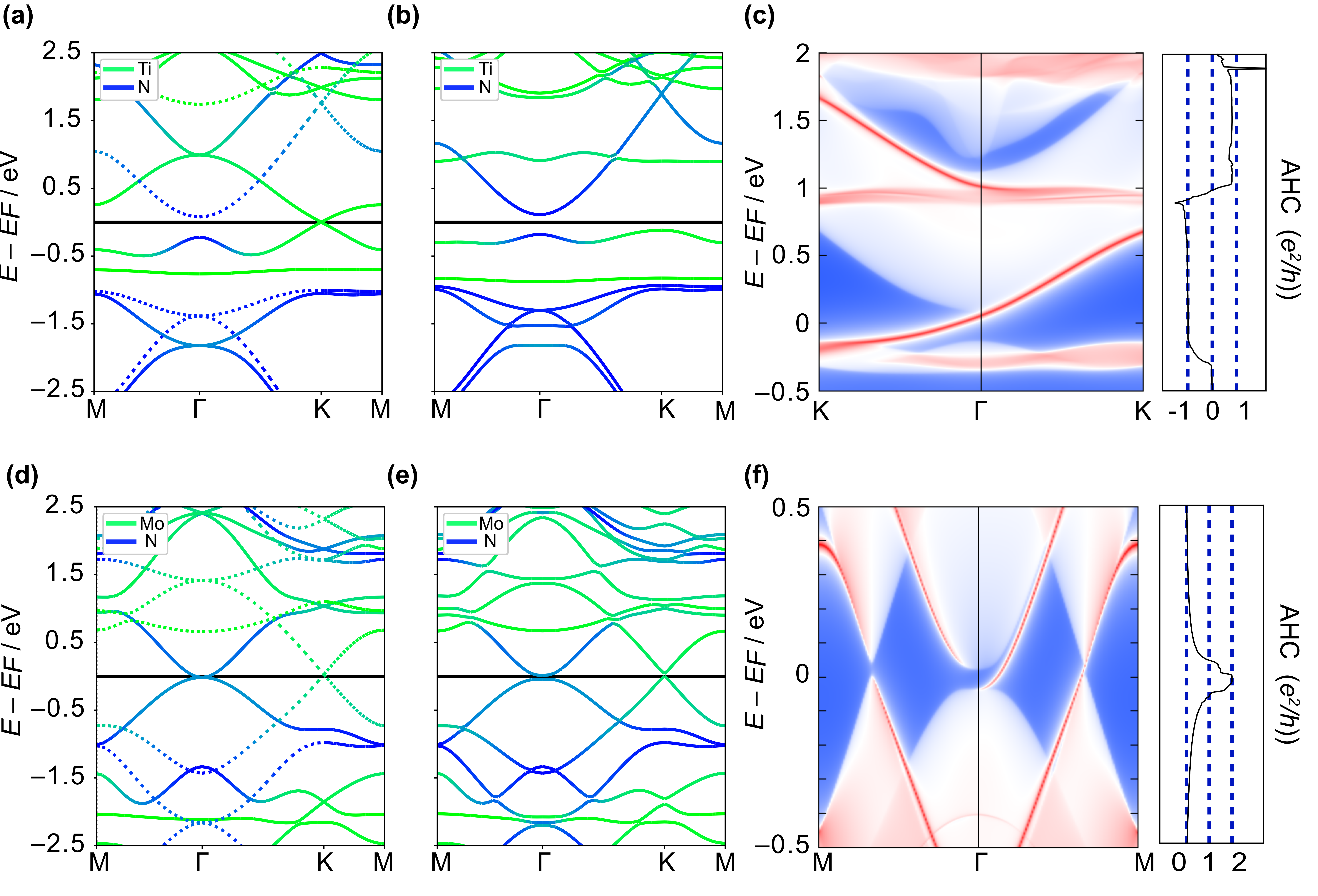}
    \caption{Element-projected band structures of (a) TiN$_8$ without SOC, (b) TiN$_8$ with SOC, (c) MoN$_8$ without SOC, and (d) MoN$_8$ with SOC. (e, f) Chiral edge states of TiN$_8$ and MoN$_8$, respectively, with the associated Chern numbers indicated on the right.}
    \label{fig7}
\end{figure*}

The spin-polarized band structures of TiN$_8$ and MoN$_8$ are presented in Fig.\ref{fig7}(a) and Fig.\ref{fig7}(d), with the dotted lines denoting the minority (spin-down) components. As discussed previously, the bands near the Fermi level are predominantly derived from Ti (Mo) $d$ orbitals. In particular, the band crossings exhibit distinct features at the $\Gamma$ and $K$ points: a parabolic shape near $\Gamma$ and a linear-like dispersion near $K$, analogous to the behavior observed in the hexagonal lattices of M$_2$X$_3$. Due to the strong localization of the $d$ electrons in MN$_8$, we classify these systems as promising candidates for topological materials.

Upon inclusion of SOC effect, the band structures (Fig.\ref{fig7}(b) for TiN$_8$ and Fig.\ref{fig7}(e) for MoN$_8$) reveal distinct SOC-induced band splittings. In TiN$_8$, a significant SOC gap (Gap 1) of 1020 meV opens at the $K$ point, accompanied by a 932 meV gap (Gap 2) at the $\Gamma$ point. In general, the indirect global band gap is 223 meV. For MoN$_8$, smaller SOC gaps of 41 meV and 63 meV are observed at the points of $K$ and $\Gamma$, respectively, which yields a global band gap of 12 meV, much smaller than that of TiN$_8$. This reduced SOC splitting is attributed to the more spatially extended 4$d$ orbitals of Mo(with Hubbard $U_{\text{Mo}} = 2\text{ eV}$), in contrast to the more localized nature of the Ti $3d$ orbitals (with Hubbard $U_{\text{Ti}} = 3\text{ eV}$). Remarkably, the SOC in TiN$_8$ leads to nearly flat bands near the Fermi level.

To further investigate the topological nature of these systems, we computed the chiral edge states and the corresponding anomalous Hall conductance (AHC), as shown in Fig.\ref{fig7}(c) and Fig.\ref{fig7}(f), respectively. In the case of TiN$_8$ (Fig.\ref{fig7}(c)), the chiral edge state traverses the Fermi level, connecting bulk $d_{xz}$ and $d_{yz}$ bands with an energy magnitude of approximately 1020 meV, but unfortunately, the bulk CBM contributed by the $p$ orbital only allows an effective band gap of 223 meV. The AHC confirms the nontrivial topological nature of this state, indicating a Chern number of $C = -1$. Additionally, a prominent SOC-induced band splitting can be observed in the energy window of 1–2 eV, where a 932 meV splitting corresponds to another nontrivial gap characterized by a Chern number close to $C = 1$. This suggests that in this system, the band crossing at $K$ and $\Gamma$ would provide Chern numbers of $-1$ and $1$ only in the spin-up subspace, respectively.

Regarding MoN$_8$, a particularly intriguing aspect lies in its unique mechanism to realize a high Chern number. As illustrated in Fig.\ref{fig7}(e), the Dirac cone-shaped band crossing at the $K$ point in MoN$_8$ originates from the spin-down subspace despite the band shapes at the $K$ and $\Gamma$ points remaining qualitatively similar to those observed in the TiN$_8$ system. This is in contrast to TiN$_8$, where the spin-up channel dominates. Fig.\ref{fig7}(f) shows the corresponding edge states, which display two chiral modes crossing the Fermi level in the same direction and a Chern number of $C = 2$ is shown on the right. Notably, the spin-up subspace in MoN$_8$ still contributes a Chern number of $C = -1$ at the $K$ point (as previously discussed). Therefore, the enhanced total Chern number in MoN$_8$ can be attributed to the reversed spin channel at the $K$ point, which produces an additional Berry curvature contribution of the same sign, effectively enhancing the topological nature. In short, in the MN$_8$ system, the band crossing contributed by $d_{xz}$ and $ d_{yz}$ orbitals could provide Chern number of $C=1$, $C=-1$ at $\Gamma$, $K$ points within the spin-up channel, respectively. For the spin-down channel, the Chern number is reversed as $C=-1$, $C=1$ at the $\Gamma$ and $K$ points.

\section{TB Hamiltonian}

Since we emphasize the TiN$_8$ motif, we carried out a detailed tight-binding Hamiltonian analysis to uncover the origin of its topology. Through fat-band analysis of TiN$_8$ in Fig.\ref{fig3}(a), we identify the band close to the Fermi level to be dominated by the Ti $d_{yz}$ that mixes with the N $p_{z}$ band except at the $\Gamma$ point. At $\Gamma$, the $p_z$ contribution is dominated by N2 sites, with only minor contributions from the N1 sites and no Ti contribution. The exchange of orbital character between the $d_{yz}$ and $p_z$-derived bands indicates an avoided-crossing behavior. This behavior can be understood from symmetry considerations. Owing to the $C_{6h}$ symmetry of TiN$_8$, the Ti site possesses the highest site symmetry $C_{6h}$, whereas the N1 and N2 sites belong to the lower-symmetry point groups $C_{3h}$ and $C_s$, respectively. At the $\Gamma$ point, the Ti $d_{xz}$ and $d_{yz}$ orbitals form the two-dimensional irreducible representation $E_{1g}$, which decomposes into $A^{\prime\prime}\oplus A^{\prime\prime}$ under $C_s$, while remaining $E^{\prime\prime}$ under $C_{3h}$. By contrast, the $p_z$ orbitals on N2 transform as $A^{\prime\prime}$ under $C_{3h}$. Consequently, the mixing between the Ti $d_{xz}/d_{yz}$ states and the N2 $p_z$ states is symmetry forbidden at $\Gamma$, whereas mixing with the lower-symmetry N1 $p_z$ states is symmetry allowed when their energies are close. Away from $\Gamma$, where the little-group symmetry is reduced, these symmetry constraints become less strict, enabling stronger orbital mixing\cite{cai2026symmetry}.

To qualitatively reproduce the topological features, we construct a simplified three-band tight-binding model with basis
($d_{xz}$, $d_{yz}$, $p^{\text{eff}}_{z}$)$^T$ to capture the electronic bands shown in Fig. S3 (a). For simplicity, we introduce an effective $p^{\text{eff}}_{z}$ orbital constructed from symmetry-adapted linear combinations (SALCs) of the N $p_z$ orbitals to describe the low-energy $p_z$-derived band. Owing to the existence of two inequivalent N environments, two distinct hexagonal motifs can be identified in the TiN$_8$ lattice (blue and cyan hexagonal loops), as illustrated in Fig. S3(b). The nearest-neighbor N1 atoms form a distorted local hexagon around each Ti atom, whereas the second-nearest-neighbor N2 atoms form a larger hexagonal network that mirrors the honeycomb plaquette structure of graphene. Due to the SALC state under the $C_{6h}$ symmetry, the effective orbital center can be assigned to the geometric center of the hexagonal plaquette (the Ti site).

The Hamiltonian can be expressed as $H=H_0 + H_{SOC}$, while the hopping term $H_0$ and the on-site SOC term $H_{SOC}$ are:
\begin{equation}
    H_0 = H_{dd}+H_{pp}+H_{dp} 
\end{equation}

\begin{equation}
    H_{SOC} = i\lambda(d^{\dagger}_{i,xz}d_{i,yz}-d^{\dagger}_{i,yz}d_{i,xz})
\end{equation}
where $H_{dd}$, $H_{dp}$, and $H_{pp}$ represent the $d_{xz}/d_{yz}$ on-site term, ${p-d}$ interaction term and ${p_{\text{eff}}}$ on-site term, respectively. $t$ and $\lambda$ denote the hopping parameters and the strength of the SOC. Finally, the Hamiltonian matrix could be described as follows
\begin{align}
\begin{gathered}
H = 
\begin{pmatrix}
\sum\limits_{\delta} t_{xz,xz} e^{i\vec{k}\cdot \vec{r}_\delta} & 
\sum\limits_{\delta} t_{xz,yz} e^{i\vec{k}\cdot \vec{r}_\delta} & 
\sum\limits_{\delta} t_{xz,p} e^{i\vec{k}\cdot \vec{r}_\delta} \\
\text{H.c.} & 
\sum\limits_{\delta} t_{yz,yz} e^{i\vec{k}\cdot \vec{r}_\delta} &
\sum\limits_{\delta} t_{yz,p} e^{i\vec{k}\cdot \vec{r}_\delta} \\
\text{H.c.} & 
\text{H.c.} &
\sum\limits_{\delta} t_{p,p} e^{i\vec{k}\cdot \vec{r}_\delta} \\
\end{pmatrix}   
  \\  +  
\begin{pmatrix}
0 & i\lambda & 0 \\
-i\lambda & 0 & 0 \\
0 & 0 & 0
\end{pmatrix}
\end{gathered}
\end{align}

\begin{figure} [htbp]
    \centering
    \includegraphics[width=0.5\textwidth]{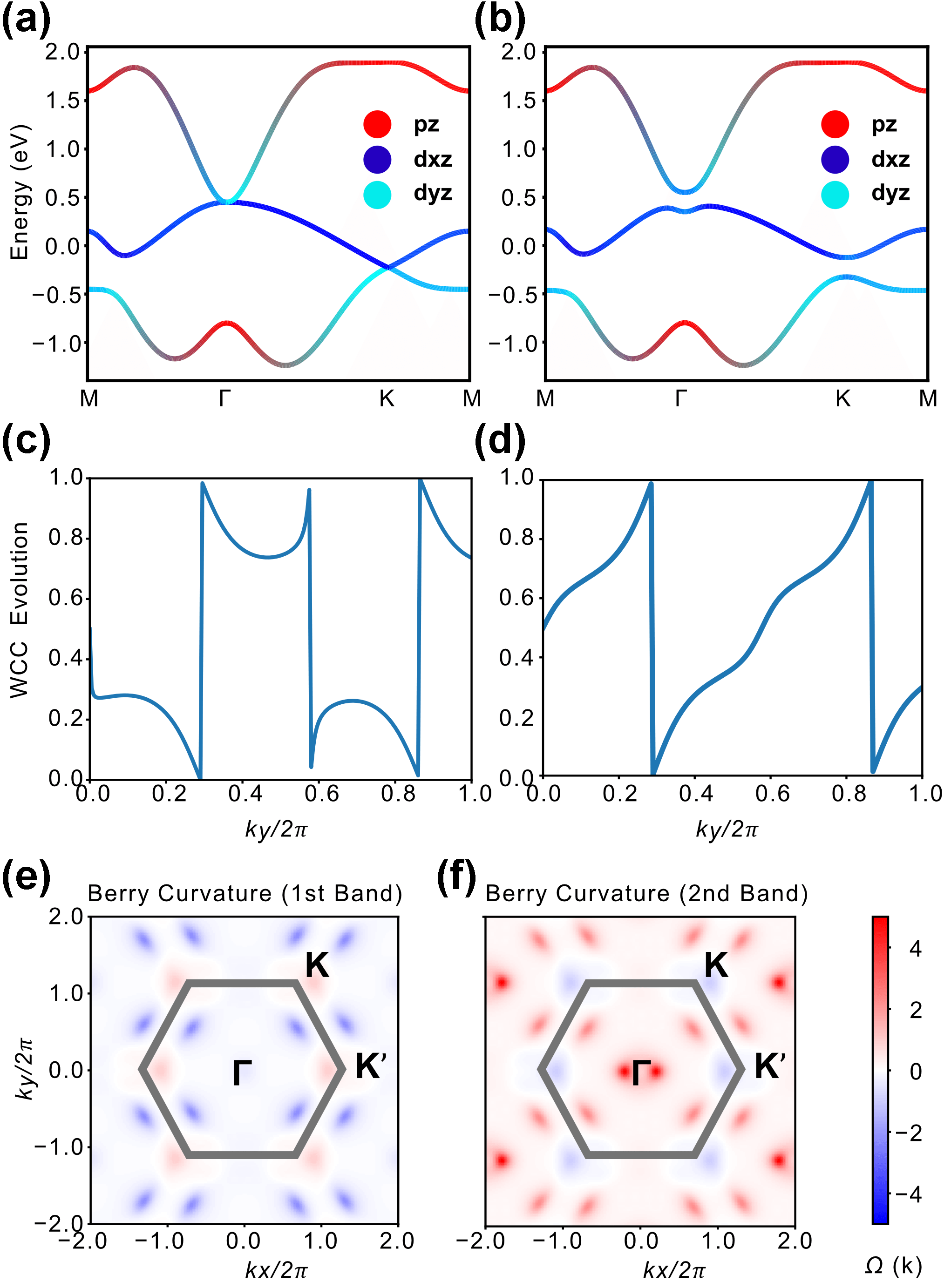}
    \caption{Simplified TB band structures for TiN$_8$. Band structures without (a) / with (b) SOC effect. Evolution of WCCs at $\Gamma$ (c) and $K$ (d). Berry curvature distribution of 1 st band (e) / 2 nd band (f) in the Brillouin zone, where the labels are counting from bottom to top in the 3 bands model.}
    \label{fig8}
\end{figure}

The hopping term $t$ is the SK parameter formulated as $t_{i,j}=<\phi _i|H_0|\phi _j>$ with details provided in the Supporting Information Table S2. Based on this Hamiltonian matrix, the calculated band structure is shown in Fig.\ref{fig8}(a) and (b). The key features at $\Gamma$ and $K$ points dominated by the $d_{xz}$ and $d_{yz}$ orbitals are well captured. When SOC is considered (Fig.~\ref{fig8} (b)), the degeneracy of the band is lifted and the magnitude of band splitting is manipulated by the SOC parameter $\lambda$. As shown in Fig.~\ref{fig8} (c) and (d), the evolution of Wannier Charge Centers (WCCs) displays the winding lines that traverse the BZ boundary along $k_y$, indicating a topologically non-trivial state. The corresponding Berry curvature distribution of the first and second bands (counting from bottom to top in the 3 bands model) in the Brillouin zone is shown in Fig.\ref{fig8}(e) and (f). The Berry curvature contributions are primarily localized around the $K$ and $K'$ points, supporting the non-trivial Chern number shown in Fig.\ref{fig8}(e). In contrast, Fig.\ref{fig8}(f) reveals an accumulation enclosing the $\Gamma$ point, while the opposite-sign contributions around $K$ and $K'$ originate from the first band. These opposite-sign contributions intuitively reflect the opposing Chern numbers associated with these two high-symmetry points.

By expanding the model around the $\Gamma$ and $K$ points, we derive two effective $k \cdot p$ models to analyze the topological contributions separately. As shown in Fig. S4(a), each massive Dirac cone at the $K$ and $K'$ valleys contributes a half-integer valley Chern number of $C=-\frac{1}{2}$, as required by symmetry, yielding a total contribution of $C=-1$ from the two valleys. Meanwhile, the $\Gamma$-centered effective model contributes a chern number of $C=1$ (Fig.S4(b)). Unlike the robust particle–hole symmetry in graphene or M$_2$X$_3$ models, the diagonal terms in the triangular lattice model are non-zero, breaking the particle–hole symmetry. And the Fermi level is determined by the occupation of electronic states, which is mediated by spin polarization.

\section{Conclusion}

In summary, by introducing magnetism and Hubbard $U$ corrections into MN$_8$ systems, we have identified a new class of QAH materials through first-principles calculations. The results demonstrate that the hypothetical TiN$_8$ material is a spin-polarized system with a lattice parameter of 5.67 Å and a robust out-of-plane magnetic anisotropy energy. Remarkably, TiN$_8$ hosts a nontrivial topological phase with a Chern number of $C=-1$ and a large topological gap of 223 meV, firmly establishing it as a robust large-gap QAH insulator. The likewise hypothetical and endothermic MoN$_8$ remains of significant interest due to its unconventional topological mechanism. Driven by the different spin channels and distinct lattice features, MoN$_8$ exhibits a high Chern number of $C=2$. However, its much smaller gap of 12 meV render it fragile and less robust than TiN$_8$. We anticipate that the exploration of emergent penta-MN$_8$ systems will advance the development of next-generation QAH materials and provide promising avenues for spintronic applications.

\section{Computational details}

First-principles calculations were performed using density functional theory (DFT) as implemented in the Vienna $ab\ initio$ Simulation Package (VASP, v 5.4.4)\cite{Kresse1993,kresseEfficientIterativeSchemes1996,KRESSE199615efficiency,kresseUltrasoftPseudopotentialsProjector1999}. Exchange and correlation were described using the generalized gradient approximation (GGA) in the Perdew-Burke-Ernzerhof (PBE) form\cite{perdewGeneralizedGradientApproximation1996}  while the pseudopotential approach followed the projector-augmented wave (PAW) method according to Blöchl\cite{blochlProjectorAugmentedwaveMethod1994}. A kinetic energy cutoff of 600 eV was used and the Brillouin zone was sampled using an $11 \times 11 \times 1$ $\Gamma$-centered $k$-mesh. The self-consistent calculations were considered converged when the total energy difference was less than 10$^{-6}$ eV, while the atomic structures were optimized until the residual Hellmann-Feynman forces were below 0.005 eV/Å. To avoid spurious interactions due to periodic boundary conditions, a vacuum width of 15 Å was added along the $z$-axis. The DFT + $U$ approach was used to account for Coulomb interactions on-site in the localized orbitals $d$, with a Hubbard $U$ value of 3.0 eV (2.0 eV) for Ti (Mo), as suggested in previous studies\cite{torres2021relevance,sakhraoui2022electronic,liu2025anomalous,liEmergenceDorbitalMagnetic2020}. Furthermore, van-der-Waals interactions were accounted for using the DFT-D3 method proposed by Grimme\cite{grimme2010consistent}. The Curie/N\'{e}el temperatures were determined using Monte Carlo simulations implemented in the VAMPIRE 7 package\cite{evans2014atomistic}, with a rectangular supercell. Each temperature step consisted of 10$^4$ equilibration steps followed by 10$^4$ measurement steps. Periodic boundary conditions were applied in the x and y directions. For structural stability, the $ab$ $initio$ molecular dynamics (AIMD) simulations were performed as $\Gamma$-only calculations with zero temperature and a time step of 2 fs under the canonical ensemble (NVT) using the Nos{\'e}-Hoover thermostat, implemented in a $4 \times 4 \times 1$ supercell. Phononcalculations were performed applying the density functional perturbation theory (DFPT) approach with a $2 \times 2 \times 1$ supercell. The Local Orbital Basis Suite Towards Electronic-Structure Reconstruction (LOBSTER, v5.1.1)\cite{dronskowski1993crystal,deringer2011crystal,maintz2013analytic,Maintz2016,nelson2020lobster,Mueller2021CrystalOrbital,Mueller2025OrbitalBased} was used to project optimized wave functions onto a local all-electron Slater-type orbital basis set (pbeVASPfit2015)\cite{Maintz2016} to obtain chemical information, such as wave function based atomic charges from Löwdin populations, calculate the COBI and DOS.

Further calculations on topological properties were based on Maximally Localized Wannier Functions (MLWFs) methods, as implemented in the Wannier90 and WannierTools packages \cite{mostofiUpdatedVersionWannier902014, wuWannierToolsOpensourceSoftware2018}. The Chern number was calculated by integrating the Berry curvature over the whole first Brillouin zone:

\begin{equation}
C=\frac{1}{2\pi}\sum_{n}\int_{\mathrm{BZ}}\ d^2k\Omega_n
\end{equation}

where $\Omega(k)$ is the Berry curvature, which is calculated according to the Kubo formula\cite{thouless1982quantized,yao2004first}:
\begin{equation}
\Omega_n\left(k\right)=-\sum_{n^\prime\neq n}\frac{2\mathrm{Im}\left\langle\psi_{nk}\left|\upsilon_x\right|\psi_{n^\prime k}\right\rangle\left\langle\psi_{n^\prime k}\left|\upsilon_y\right|\psi_{nk}\right\rangle}{\left(\varepsilon_{n^\prime}-\epsilon_n\right)^2}
\end{equation}

\textbf{Supporting Information}  
Supporting Information is available from the Wiley Online Library or from the author.


\textbf{Acknowledgments} 
This work was supported by the GuangDong Basic and Applied Basic Research Foundation (Grant No. 2023A1515110719).

\textbf{Conflict of Interest}
The authors declare no conflict of interest.


%
%




\end{document}



\title{SI for "Large-gap quantum anomalous Hall insulator in two-dimensional pentagonal TiN\(_{8}\) monolayer"}

\author{Zheng Wang$^{1,\#}$, Linda Sophie Reitz$^{4,\#}$, Beichen Ruan$^{1}$, Zhuoheng Li$^{1}$, Shu-Shen Lyu$^{1-3}$, Richard Dronskowski$^{4,*}$, Kaixuan Chen$^{1-3,*}$}

\email{chenkx26@mail.sysu.edu.cn; drons@HAL9000.ac.rwth-aachen.de}

\affiliation
{$^1$School of Materials, Shenzhen Campus of Sun Yat-sen University, Shenzhen, 518107, PR China, $^2$Guangdong Engineering Technology Research Centre for Advanced Thermal Control Material and System Integration (ATCMSI), Sun Yat-sen University, Guangzhou, 510275, PR China, $^3$Huizhou Research Institute, Sun Yat-sen University, Huizhou, 516081, PR China, $^4$Chair of Solid-State and Quantum Chemistry, Institute of Inorganic Chemistry, RWTH Aachen University, D-52056 Aachen, Germany. $^{\#}$These authors contribute equally to this work.}


\date{\today}

\maketitle

\section{Complement to the Hubbard $U$ test}
We performed a rough scan of $U$ values with a step of 0.5 eV for both systems, to investigate the Hubbard $U$ values-dependent magnetism. The corresponding energy differences are shown in Fig.S1 and Table.2 in maintext. 


\begin{figure} [h]
    \centering
    \includegraphics[width=0.5\textwidth]{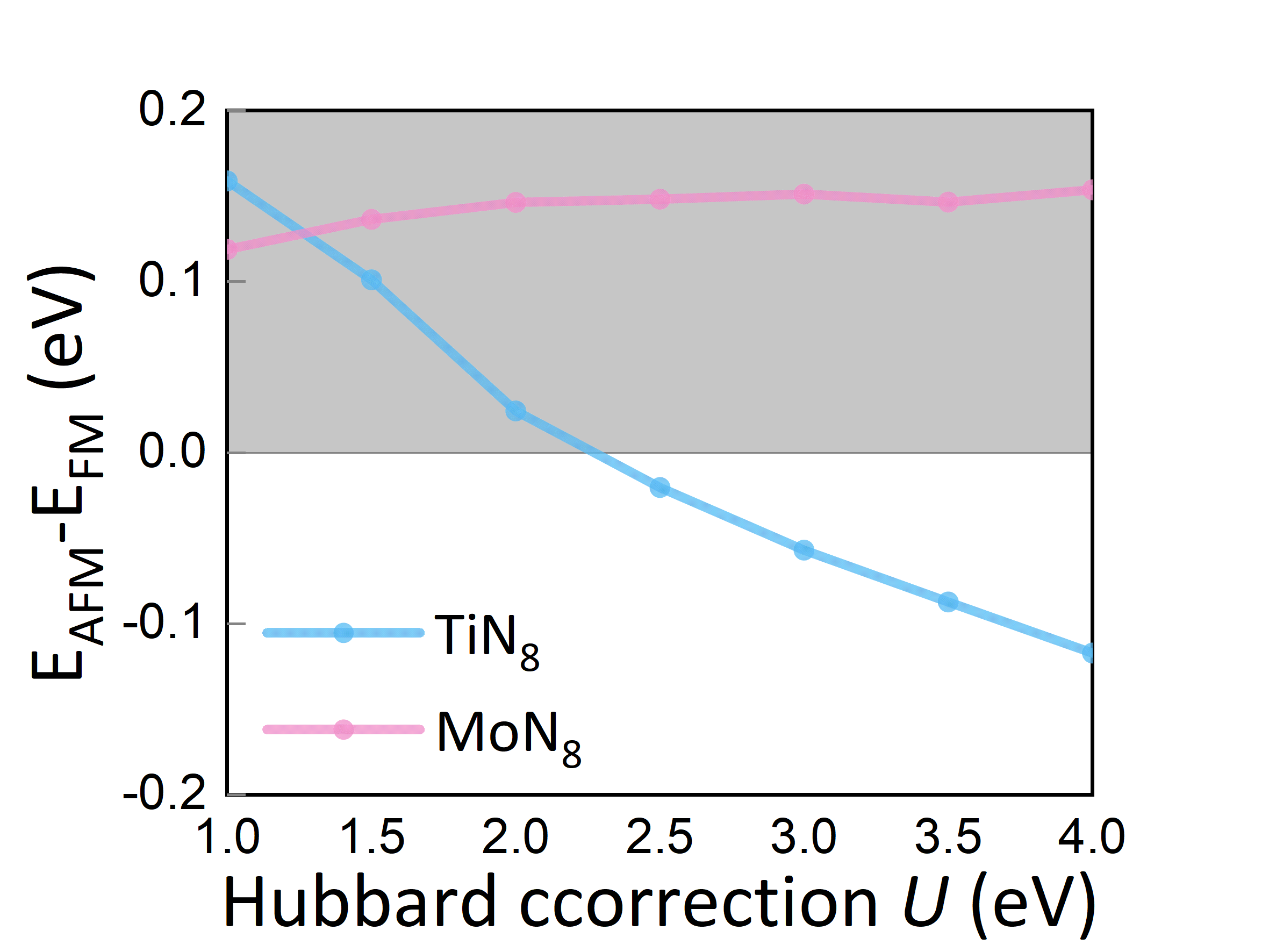}
    \caption{Energy differences $\Delta E$ (defined as $E_{AFM}-E_{FM}$) for both systems within 2$\times$2$\times$1 supercell.}
    \label{figs.1}
\end{figure}

\section{Complement to the AIMD simulation of MoN$_8$}

\begin{figure} [h]
    \centering
    \includegraphics[width=0.8\textwidth]{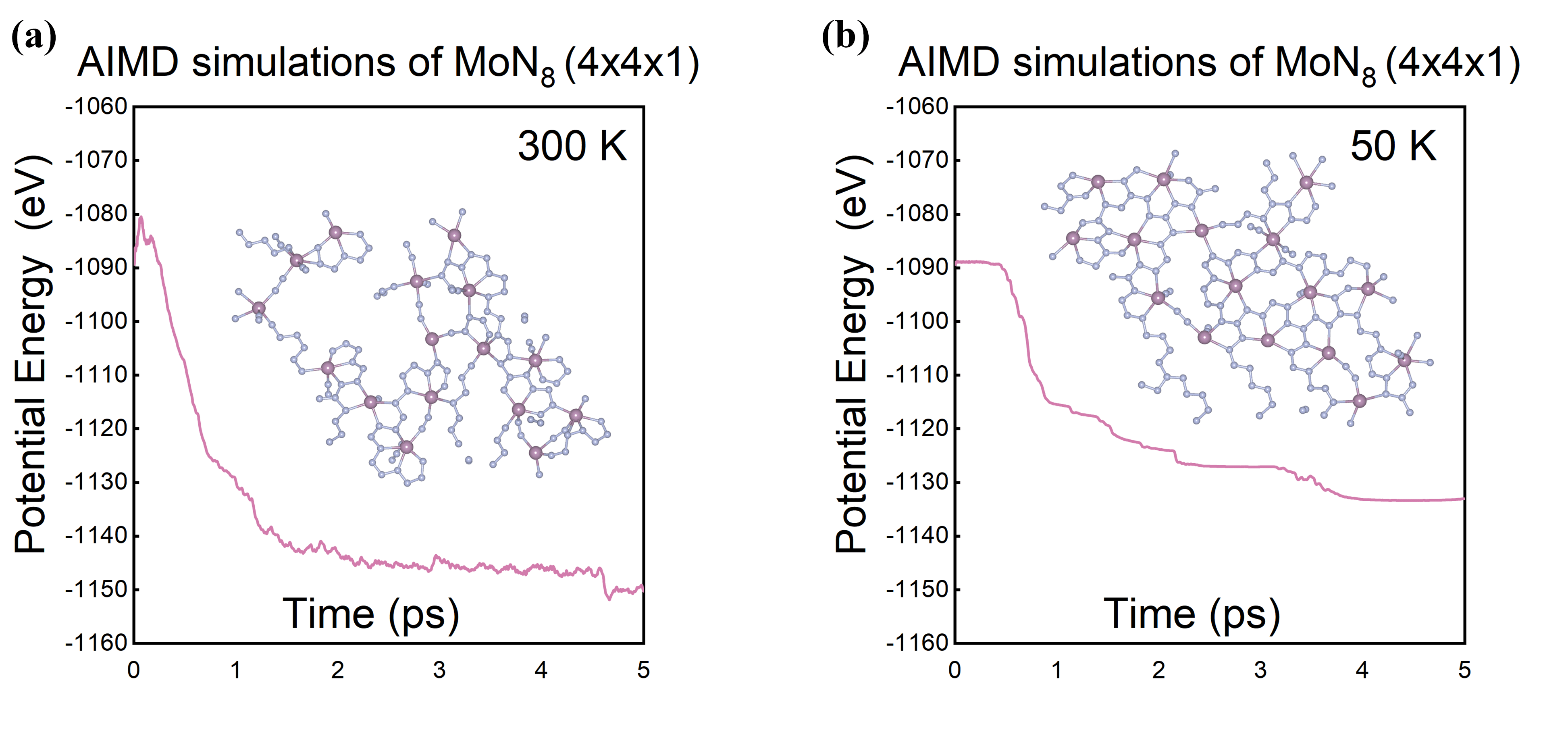}
    \caption{AIMD  potential energy results of MoN$_8$ with snapshots at (a) 300 K and (b) 50 K, within 4$\times$4$\times$1 supercell.}
    \label{figs.2}
\end{figure}

\section{Complement to the total energies in different magnetic configurations of both systems}
\begin{table}[htbp]
\centering
\caption{Total energies of magnetic configurations}
\label{tab:my-tables2}
\begin{tabular*}{\textwidth}{@{\extracolsep{\fill}} lcccccc @{}}  
\hline\hline
Structures & \multicolumn{3}{c}{TiN$_8$} & \multicolumn{3}{c}{MoN$_8$} \\ 
\hline
Functionals & PBE+$U$ & PBE & SCAN & PBE+$U$ & PBE & SCAN \\ \hline
FM (eV) & -266.57 & -272.89 & -337.24 & -265.53 & -272.37 & -398.27 \\
AFM-S (eV) & -266.77 & -272.66 & -336.92 & -265.39 & -272.26 & -398.13 \\
AFM-Z (eV) & -266.78 & -272.63 & -336.88 & -265.27 & -272.29 & -398.07 \\ 
\hline\hline
\end{tabular*}
\end{table}

\section{Simplified TB Hamiltonian and Chern number}

\[
\vec{e}_1 = (a,0), \;
\vec{e}_2 = \left(\tfrac{a}{2}, -\tfrac{\sqrt{3}\,a}{2}\right), \;
\vec{e}_3 = \left(-\tfrac{a}{2}, -\tfrac{\sqrt{3}\,a}{2}\right), \;
\vec{e}_4 = (-a,0), \;
\vec{e}_5 = \left(-\tfrac{a}{2}, \tfrac{\sqrt{3}\,a}{2}\right), \;
\vec{e}_6 = \left(\tfrac{a}{2}, \tfrac{\sqrt{3}\,a}{2}\right).
\]

\begin{figure} [htbp]
    \centering
    \includegraphics[width=0.8\linewidth]{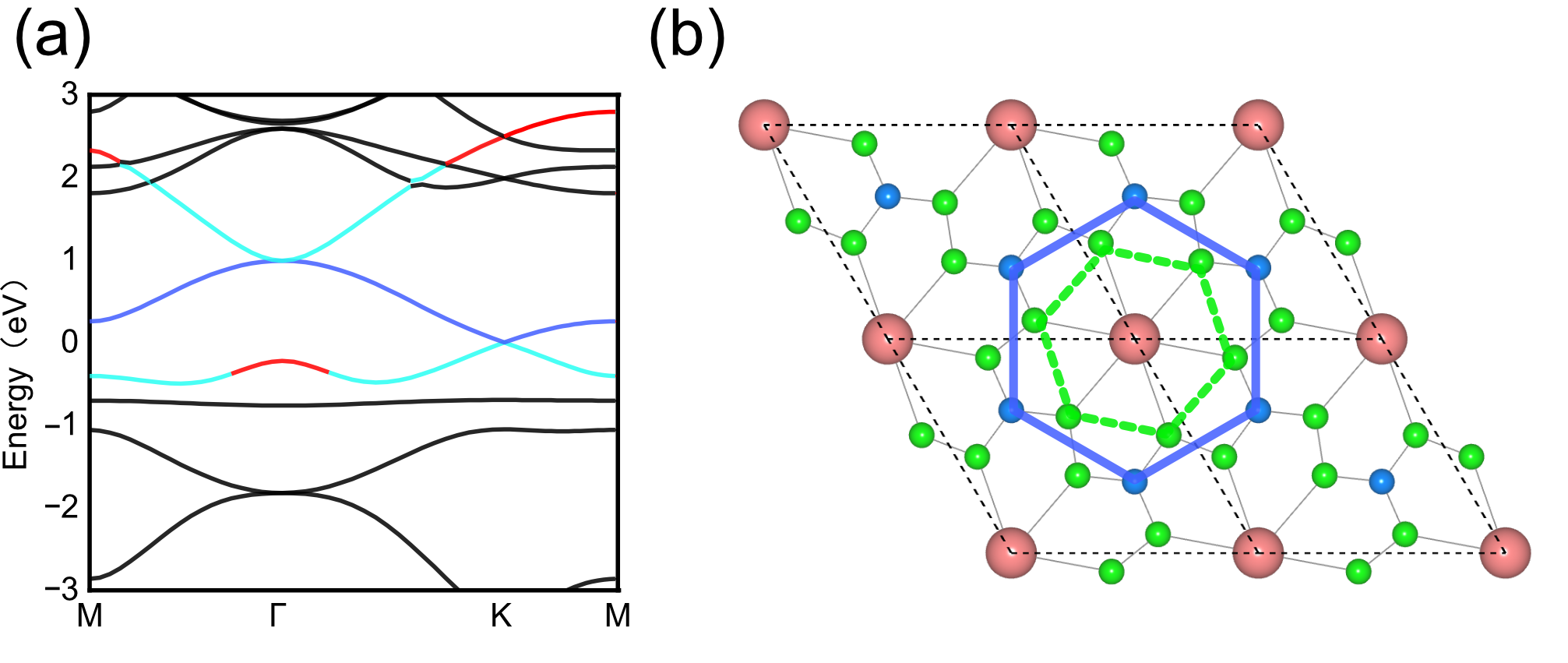}
    \caption{(a) The chosen relevant 3 bands projected from DFT result of TiN$_8$. (b) The two inequivalent N environments and two distinct hexagonal loops, where N1 (N2) sites are denoted in green (blue).}
    \label{figs3}
\end{figure}

\subsection{TB matrix and Chern number around $\Gamma$}

At $\Gamma$, the expansion could be expressed as:
\begin{equation}
H_\text{eff}({\Gamma})= \begin{pmatrix}
|\vec{\mathbf{q}}|^{2}-q_{x}^{2} & -q_{x}q_{y}+i\lambda \\
-q_{x}q_{y}-i\lambda & |\vec{\mathbf{q}}|^{2}-q_{y}^{2}
\end{pmatrix}
\end{equation}
Apparently, when SOC is excluded, this expression yields a 
quadratic (parabolic)  dispersion near the $\Gamma$ point : $E\propto|\mathbf{q}|^2$. In analogy to other materials with normal rotational symmetry, this leads to a conventional second-order band shape. However, the inclusion of SOC introduces a nontrivial band splitting at $\Gamma$, reducing the degeneracy.

Around the $\Gamma$ point ($\vec{k} \to \vec{q} \approx 0$), it gives
\begin{align}
H_{11} &= 2t \cos(\vec{k}\cdot\vec{e}_1) 
+ \frac{t}{2}\left[\cos(\vec{k}\cdot\vec{e}_2) + \cos(\vec{k}\cdot\vec{e}_3)\right] \nonumber \\
&\approx 2t\left(1 - \frac{q_x^2}{2}\right)
+ \frac{t}{2} \left[2 - \frac{q_x^2}{4} - \frac{3 q_y^2}{4}\right] \nonumber \\
&= 3t - \frac{9 q_x^2 + 3 q_y^2}{8}t.
\end{align}

Similarly, the other matrix elements read:
\begin{align}
H_{12} &= - \frac{3 q_x q_y}{4}t + i\lambda, \\
H_{21} &= H_{12}^*, \\
H_{22} &= 3t - \frac{3 q_x^2 + 9 q_y^2}{8}t.
\end{align}

The general two-band Hamiltonian is the following
\begin{equation}
    H(\mathbf{k})=\epsilon_{0}(\mathbf{k})\mathbb{I}+\textbf{d}(\mathbf{k})\vec{\sigma}
\end{equation}

Therefore, the vector $\textbf{d}(\textbf{q})$ reads
\begin{equation}
\textbf{d}(\textbf{q}) =
\left(
-\frac{3 q_x q_y}{4}t,\ \lambda,\ -\frac{3 q_x^2 - q_y^2}{8}t
\right)^T
\end{equation}

Thus, the Berry curvature can be written as follows.
\begin{align}
\Omega(\mathbf{q})
&= \frac{1}{2|\mathbf{d}(\mathbf{q})|^{3}} 
\mathbf{d}(\mathbf{q}) \cdot 
\left( \partial_{k_x}\mathbf{d}(\mathbf{q}) \times \partial_{k_y}\mathbf{d}(\mathbf{q}) \right) \notag \\
&= \frac{144 \lambda q^{2}}{\left( 9 q^{4} + 64 \lambda^{2} \right)^{3/2}} .
\end{align}

The Chern number is obtained by integrating the Berry curvature over the entire momentum space.
\begin{align}
C &= \frac{1}{2\pi} \iint_{\mathbb{R}^2} \Omega(\mathbf{q}) \, d^{2}\mathbf{q} \notag \\
&= \frac{1}{2\pi} \int_{0}^{2\pi} d\phi \int_{0}^{\infty} 
\frac{18 \lambda q^{2}}{\left( 9 q^{4} + 64 \lambda^{2} \right)^{3/2}} \, q \, dq \notag \\
&= \frac{\lambda}{|\lambda|} = 1 .
\end{align}

\subsection{TB matrix and Chern number around $K$}

The $k \cdot p$ expansion around $K$ is:

\begin{equation}
H_\text{eff}(K)=
\begin{pmatrix}
q_x & -q_y+i\lambda \\
-q_y-i\lambda & -q_x
\end{pmatrix}
\end{equation}
Where  ${\vec{\textbf{q}}}=\vec{k} \space -K $. 
When SOC is not included ($\lambda=0$), the result gives a linear dispersion at the point $K$: $E\propto|\vec{\mathbf{q}}|$. All elements of the effective Hamiltonian $H_0$ are real, distinguishing this model from the graphene and M$_2$X$_3$ cases, although both exhibit linear dispersion at $K$. This indicates the emergence of an unconventional Dirac point that is driven by the sublattice symmetry. At the $K$ point $\left(\frac{2\pi}{3},\frac{2\pi}{\sqrt{3}}\right)$, the Hamiltonian reads:
\begin{align}
H_{11} &= \left(\frac{3}{2}+\frac{3\sqrt{3}\,q_x}{4}\right) t, \\
H_{12} &= - \frac{3\sqrt{3}\,q_y}{4}t - i\lambda, \\
H_{21} &= H_{12}^*, \\
H_{22} &= \left(\frac{3}{2}-\frac{3\sqrt{3}\,q_x}{4}\right) t.
\end{align}
The corresponding Berry curvature calculation yields the Chern number:
\begin{equation}
C = -\frac{\lambda}{2|\lambda|} = -\frac{1}{2}.
\end{equation}

\begin{figure} [htbp]
    \centering
    \includegraphics[width=0.5\linewidth]{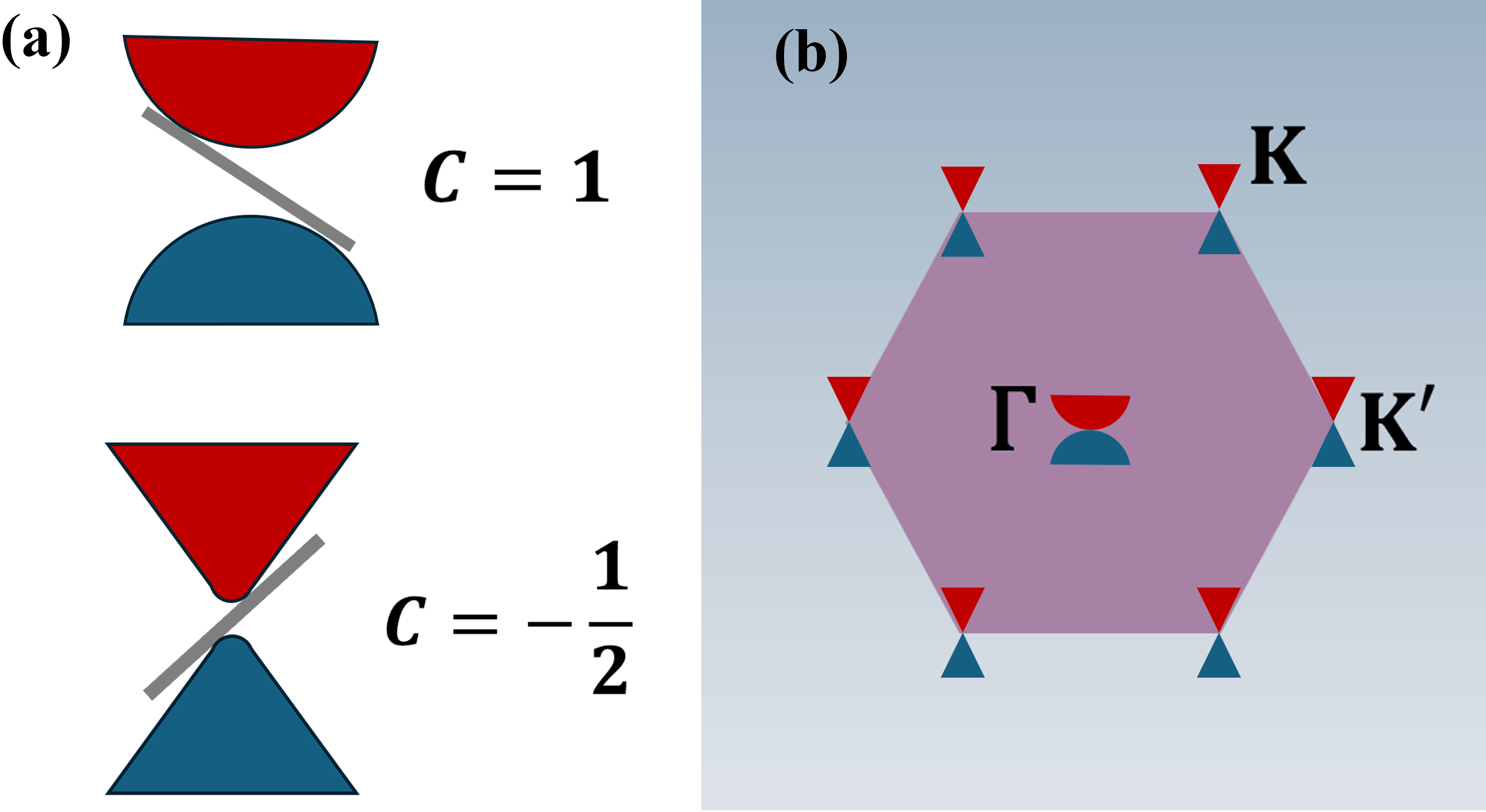}
    \caption{Schematic illustration of topologically nontrivial band gaps. (a) Comparison between non-Dirac-like and Dirac-like band dispersions and their associated Chern numbers. (b) The position of the band-crossing points in the Brillouin zone.}
    \label{figs4}
\end{figure}
